# 1.2 dB Insertion Loss in a Monolithically Integrated Silicon Mach–Zehnder Optical Isolator Using Stoichiometric Ce:YIG

ZHENYUAN REN[1,2,†], DI WU[1,2,†], TIANCHI ZHANG[1,2], YIZHI CHEN[1,2], JUNXIAN WANG[1,2], ZIXUAN WEI[1,2], JUN QIN[1,2], LEI BI[1,2,*]

[1]*National Engineering Research Center of Electromagnetic Radiation Control Materials, University of Electronic Science and Technology of China, Chengdu 611731, China.*
[2]*Key Laboratory of Multi-spectral Absorbing Materials and Structures of Ministry of Education, University of Electronic Science and Technology of China, Chengdu 611731, China.*
[†]*These authors contributed equally.*
**E-mail: bilei@uestc.edu.cn*

**Abstract:** Integrated optical isolators play a crucial role in photonic integrated circuits (PICs) by effectively stabilizing optical signals, making them indispensable components for advanced photonic systems. However, integrated magneto-optical (MO) isolators suffer from their high insertion loss, which is mainly caused by the relatively low FoM of Ce:YIG MO thin films deposited on silicon. Here, we show that film stoichiometry plays a key role in controlling the FoM of Ce:YIG. We observed garnet phase Fe-rich and Ce-rich nano-clusters in non-stoichiometric Ce:YIG. By using an iron-deficient target, we demonstrated that nearly stoichiometric Ce:YIG films on Si(100) could be fabricated, eliminating the nano-cluster defects. Consequently, the film exhibited a saturation Faraday rotation angle of -5523 deg/cm and a figure of merit of 72.7 deg/dB, which is 1.7 times higher compared to nonstoichiometric films. An MZI-type optical isolator on SiN fabricated by this method demonstrated a record low insertion loss of 1.2 dB and isolation ratio of 27.6 dB at 1556 nm wavelength. Our results provide an important step toward the practical application of integrated MO nonreciprocal photonic devices.

## 1. Introduction

The rapid advancement of artificial intelligence (AI) has led to an ever-increasing demand for computational resources and digital hardware performance [1]. Photonic integrated circuits (PICs) have been widely used in data communication due to their unique advantages such as high bandwidth and low energy consumption [2]. Among key components of PICs, nonreciprocal photonic devices such as optical isolators have been extensively studied for their ability to break time-reversal symmetry in light propagation, which is essential for blocking reflection into lasers, amplifiers, photonic subsystems, as well as full optical communication [3]. Three main approaches have been applied for achieving optical isolation based on nonlinear optical effects [4-7], spatio-temporal modulation [8-10], and the magneto-optical (MO) effect [11, 12]. MO optical isolators show advantages of passive device operation, simple device structure, and broadband. A variety of MO nonreciprocal photonic devices have been demonstrated based on micro-ring resonators, multimode interferometers, and Mach-Zehnder interferometers [13-17].

However, silicon integrated MO isolators suffer from relatively high insertion loss. Compared to commercial bulk devices with insertion loss of less than 0.5 dB, integrated optical isolators typically show insertion loss up to 2.3~10 dB [14, 16-23]. Although ring-resonator-type optical isolators showed lower loss, they suffered from narrow operation bandwidth which require complex feedback control during application [16, 18]. To reduce the insertion loss, research activities primarily undertook two approaches: device design and material optimization. Losses originating from the device itself include waveguide propagation loss, bending loss and junction loss. Waveguide propagation loss could be reduced by using low loss waveguide materials, such as SiN [16]. Bending loss and junction loss could be reduced by optimizing device design [14, 16, 24-26]. The remaining losses were attributed to the MO material layer, specifically, the garnet films. Currently, the FoM of Ce:YIG thin films deposited on SiN has been reported to be 42 deg/dB [16], resulting in a 2.3 dB insertion loss in an MZI-type optical isolators. Compared to epitaxial Ce:YIG films with FoM up to 943 deg/dB, a clear gap exists between the material quality [27].

In this work, we demonstrated that Ce:YIG film stoichiometry plays a key role in controlling the material loss and FoM. Due to the preferential ablation of Fe during pulsed laser deposition, we demonstrated that nearly stoichiometric $Ce_{1.4}Y_{1.6}Fe_{4.89}O_{12}$ MO thin films by employing an iron-deficient Ce:YIG target. Compared to films deposited using a stoichiometric target, the Ce:YIG films grown from the iron-deficient target exhibited fewer defects such as Fe-rich and Ce-rich precipitates and improved crystalline quality. The $Ce_{1.4}Y_{1.6}Fe_{4.89}O_{12}$ film deposited on silicon showed Faraday rotation of -5523 deg/cm at 1550 nm, and FoM of 72.7 deg/dB, 73.1 % higher than non-stoichiometric films. The fabricated MZI type optical isolator showed a record low insertion loss of 1.2 dB with an isolation ratio of 27.6 dB at a wavelength of 1556 nm.

## 2. Methods

### *2.1 MO Thin Film Fabrication and Characterization*

Polycrystalline Ce:YIG films on Si were prepared using a two-step deposition method. First, a 55 nm thick YIG seed layer was deposited on Si(100) substrate by sputtering at room temperature and subsequently crystallized by annealing in a rapid thermal annealing (RTA) furnace under an oxygen pressure of 2 Torr for 300 seconds, followed by natural cooling [28]. Then, the Ce:YIG films were deposited by pulsed laser deposition (PLD) at 750 °C. A Compex Pro 205 KrF excimer laser ($\lambda$ = 248 nm, repetition rate = 10 Hz) was employed for film deposition. The target-to-substrate distance was set at 56 mm, and the base vacuum pressure was $1 \times 10^{-6}$ mbar. Ceramic targets were prepared using a standard solid-state reaction method with high-purity $Y_2O_3$ (99.99%), $CeO_2$ (99.99%), and $Fe_2O_3$ (99.945%) powders as starting materials. After deposition, the sample was cooled to room temperature at 5 °C/min.

To investigate the effect of different Fe concentrations on the structure and properties of Ce:YIG films, the following characterizations were conducted. X-ray diffraction (XRD) was used to determine the phase composition of the Ce:YIG films. $\omega$–2$\theta$ scans were conducted on an Ultima IV diffractometer with Cu-$K_\alpha$ radiation ($\lambda$ = 1.54 Å), spanning a 2$\theta$ range of 25°–37° at a step size of 0.02°. Atomic force microscopy (AFM) was employed to observe the surface morphology and roughness of the films. The AFM images were acquired in tapping mode using gold-coated silicon probes with a resonant frequency of 300 kHz. High-resolution transmission electron microscopy (HRTEM) was employed to characterize the crystal structure and defects of the films. To further determine the chemical composition of these defects, energy-dispersive X-ray spectroscopy (EDS) mapping was conducted on the films. Electron backscatter diffraction (EBSD) was employed to characterize the crystalline quality of the films. To analyze the valence states of Ce ions, X-ray photoelectron spectroscopy (XPS) was performed using monochromatic Al $K_\alpha$ radiation on an Axis Supra system. Inductively coupled plasma optical emission spectrometry (ICP-OES) with a Perkin Elmer Optima 8000 instrument was employed to determine the elemental composition (Fe, Ce, and Y) of the Ce:YIG films. Magnetic hysteresis loops were measured at room temperature by vibrating sample magnetometry (VSM) using a Cryogenic (UK) system equipped with a superconducting magnet. The Faraday rotation hysteresis loops were measured using a home-built Faraday rotation angle measurement system. In the following, we refer to the films by their actual stoichiometries.

### *2.2 Optical Isolator Fabrication and Characterization*

The proposed devices were first fabricated in a silicon photonics foundry. The SiN waveguide was fabricated on a $SiO_2$ bottom cladding using low pressure chemical vapor deposition (LPCVD) followed by reactive ion etching (RIE). Subsequently, chemical mechanical polishing (CMP) was employed to planarize the top $SiO_2$ cladding layer, after which RIE was used to etch and expose the SiN waveguide cores in the MO waveguide regions. Finally, magnetron sputtering and PLD were utilized to deposit 55 nm thick YIG films and 130 nm thick Ce:YIG films, respectively.

The devices were characterized on a polarization-maintaining, fiber-butt-coupled system. During the measurements, an external in-plane magnetic field of 2000 Oe was applied perpendicular to the light propagation direction in the MO/SiN waveguides to fully saturate the garnet films. A tunable laser provided a continuous wave spectrum spanning from 1520 nm to 1610 nm, with the input optical power set at 8 dBm. A free-space polarization controller was employed to generate the desired TE or TM polarized light. The optical signal was coupled into the device under test via a lensed PM fiber, and the output signal was subsequently coupled into another fiber connected to an optical power sensor. The wavelength detection interval was set to 2 pm. For backward transmission characterization, the direction of the applied magnetic field was kept fixed. The propagation direction of incident laser was reversed.

## 3. Results and discussion

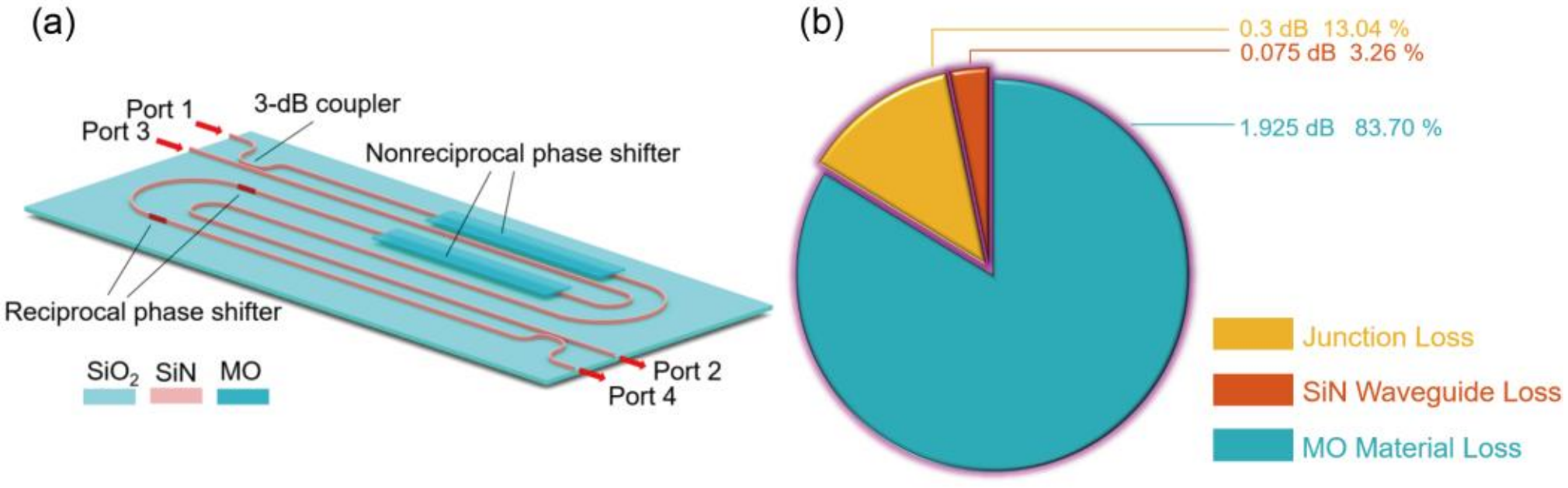

Fig. 1. (a) Schematics of on-chip MZI type optical isolator based on Ce:YIG magneto-optical thin films on SiN. (b) Quantitative loss breakdown diagram in the isolator devices.

Figure 1(a) schematically illustrated the SiN-based optical isolator. Operating on the fundamental principle of Mach-Zehnder interference, the device dictates optical transmission by manipulating the phase difference between its two arms. Specifically, constructive interference allows forward-propagating light to pass, while destructive interference effectively blocks backward-propagating light. Structurally, the architecture comprises two 3 dB directional couplers, two reciprocal phase shifters (RPS), and two nonreciprocal phase shifters (NRPS). To minimize the footprint, both interference arms of the MZI were routed on the same side, yielding a compact device area of $1.1 \times 0.24$ $mm^2$ excluding the polarization rotator. Integrating 3 dB directional couplers at both ends established a four-port circulator architecture. Here, the reciprocal and nonreciprocal phase shifters (NRPS) jointly provide a 0 ($\pi$) phase difference for forward (backward) propagating light. Specifically, the NRPS relies on MO waveguide—comprising Ce:YIG/YIG thin films on SiN channel waveguides—to induce the phase shift for TM-polarized light. Furthermore, in racetrack resonator-based narrow-band isolators, the MO waveguide was positioned outside the bus-ring coupling region, achieving optical isolation via resonance frequency splitting driven by an intracavity NRPS.

Figure 1(b) presented the quantitative loss breakdown of our previously demonstrated SiN-based MO MZI isolator [16]. Showing a total insertion loss of 2.3 dB, this device represented the state-of-the-art performance for MZI-type MO isolators. The total loss comprised three components: SiN waveguide loss, junction loss, and MO material loss. Benefiting from the intrinsically low-loss SiN platform and a highly optimized window-opening fabrication process, the SiN waveguide loss was reduced to 0.075 dB (accounting for 3.26% of the total), and the junction loss between MO and non-MO waveguides was reduced to 0.3 dB (13.04%). Consequently, the remaining 1.925 dB was attributed to the MO material, dominating 83.70% of the overall insertion loss. This substantial material penalty was fundamentally attributed to the incongruent elemental transfer during the PLD process. The consequent stoichiometric deviation in the deposited Ce:YIG films triggered the formation of secondary-phase precipitates, which exacerbated optical scattering and absorption. Therefore, a systematic investigation and precise control of the Ce:YIG film stoichiometry were imperative to fundamentally suppress these defects and further drive down the insertion loss of integrated MZI devices.

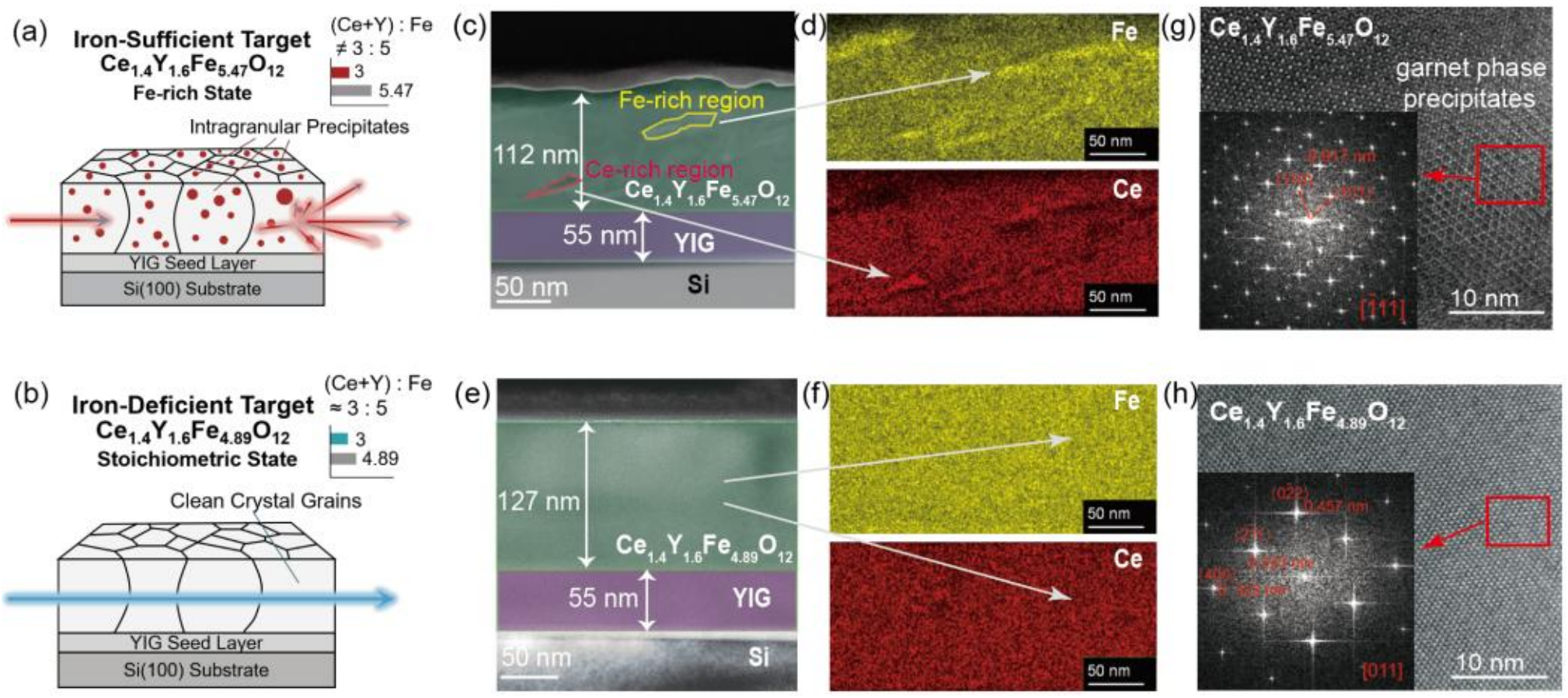

Fig. 2. Microstructural comparison and elemental analysis of the Ce:YIG films. Schematic comparison between films deposited from (a) iron-sufficient and (b) iron-deficient targets. (c) Cross-sectional TEM image of the off-stoichiometric $Ce_{1.4}Y_{1.6}Fe_{5.47}O_{12}$ film. (d) EDS elemental map of Fe element and Ce element of the $Ce_{1.4}Y_{1.6}Fe_{5.47}O_{12}$ film. (e) Cross-sectional TEM image of the nearly stoichiometric $Ce_{1.4}Y_{1.6}Fe_{4.89}O_{12}$ film. (f) EDS elemental map of Fe element and Ce element of the $Ce_{1.4}Y_{1.6}Fe_{4.89}O_{12}$ film. (g) High-resolution TEM image of the precipitation in the $Ce_{1.4}Y_{1.6}Fe_{5.47}O_{12}$ film. (h) High-resolution TEM image of the Ce:YIG region.

To investigate the impact of film stoichiometry on the material's microstructure, comprehensive material characterizations were performed. Figure 2(a) illustrated the microstructure of an iron rich Ce:YIG film deposited using a stoichiometric target, while Figure 2(b) displayed the Ce:YIG film grown from an iron-deficient target, which revealed impact of stoichiometry on the microstructure and optical performance of Ce:YIG films. The Fe-rich state induced precipitates that caused severe optical loss. In contrast, the nearly stoichiometric $Ce_{1.4}Y_{1.6}Fe_{4.89}O_{12}$ films with clean crystal grains, effectively minimizing optical loss and enhancing the MO figure of merit. During the PLD process, the ablation rate of Fe atoms generally exceeds that of Ce and Y elements [29-31]. Consequently, utilizing an iron-deficient target compensated for this elemental loss, yielding films with nearly stoichiometric ratio. This hypothesis was confirmed by ICP-OES measurement, which revealed a film composition (Ce+Y) : Fe = 3 : 4.89 from the iron-deficient target and 3 : 5.47 from the stoichiometric target. Under identical deposition conditions, the resulting $Ce_{1.4}Y_{1.6}Fe_{5.47}O_{12}$ films exhibited considerable microstructural defects, primarily characterized by widespread Fe-rich and Ce-rich precipitates localized within the garnet crystal grains. In sharp contrast, the $Ce_{1.4}Y_{1.6}Fe_{4.89}O_{12}$ films deposited from the iron-deficient target exhibited a cleaner microstructure. The formation of precipitates was significantly suppressed, leaving the regions structurally clean and defect-free. This is clearly evidenced by the wide-field TEM images shown in Figures S3(a) and (b).

Figure 2(c) displayed the cross-sectional TEM image of the off-stoichiometric $Ce_{1.4}Y_{1.6}Fe_{5.47}O_{12}$ film, which demonstrated non-uniformity and structural defects within the crystalline grains. EDS was utilized to reveal the chemical composition of the defects in Figure 2(d). The elemental maps of Fe and Ce for the $Ce_{1.4}Y_{1.6}Fe_{5.47}O_{12}$ film in Figure 2(d) revealed spatial phase segregation, where Fe-rich and Ce-rich domains were adjacent to one another. Supplementary Figure S3(c) provided the complete elemental mappings, showing that both the Ce-rich and Fe-rich regions contained O elements while lacking Y elements. This result proved that the defect regions were oxides of Ce or Fe. In sharp contrast, the nearly stoichiometric $Ce_{1.4}Y_{1.6}Fe_{4.89}O_{12}$ film exhibited a uniform and structurally clean morphology within the crystal grains, as observed in Figure 2(e). High-resolution TEM (HRTEM) of this region revealed highly ordered periodic lattice fringes, indicating excellent crystalline quality. Figure 2(f) presented the EDS elemental mappings of the $Ce_{1.4}Y_{1.6}Fe_{4.89}O_{12}$ film, demonstrating the uniform distribution of Fe and Ce elements. This microstructural distinction was further corroborated by Supplementary Figure S2. EBSD analysis revealed that the $Ce_{1.4}Y_{1.6}Fe_{4.89}O_{12}$ film exhibited a high garnet phase purity of 98.9%. In contrast, the phase purity of the $Ce_{1.4}Y_{1.6}Fe_{5.47}O_{12}$ film degraded to 94.1% due to the formation of Ce or Fe-rich precipitates. Furthermore, AFM characterizations demonstrated that despite exhibiting comparable surface roughness ($R_q$ = 5.46 nm for $Ce_{1.4}Y_{1.6}Fe_{4.89}O_{12}$ and $R_q$ = 5.18 nm for $Ce_{1.4}Y_{1.6}Fe_{5.47}O_{12}$).

Figures 2(g) and (h) showed HRTEM images of the defect region and the Ce:YIG matrix region, respectively. The fast Fourier transform (FFT) pattern in the inset of Figure 2(g), obtained from the Fe-rich region, was consistent with the $[\bar{1}11]$ zone axis. The reflections indexed as (110) and (101) corresponded to an interplanar spacing of 0.917 nm, which was consistent with the garnet phase. Together, the EDS elemental maps and HRTEM images showed that the off-stoichiometric $Ce_{1.4}Y_{1.6}Fe_{5.47}O_{12}$ film contained Ce- and Fe-rich defect regions, with garnet-phase precipitates observed in these regions. The FFT pattern in the inset of Figure 2(h) was indexed along the [011] zone axis. The measured interplanar spacings of 0.457, 0.523, and 0.322 nm were assigned to the $(0\bar{2}2)$, $(2\bar{1}1)$, (400) planes, respectively. These indices and spacings were consistent with the garnet structure of the Ce:YIG matrix. By employing an iron-deficient target, we successfully suppressed the formation of these precipitates.

Figure 3 presented the room temperature magnetic hysteresis loops and free-space Faraday rotation of the Ce:YIG films with varying iron content. Figure 3(a) and 3(b) showed the in-plane and out-of-plane magnetic hysteresis loops, respectively. All films exhibited easy-plane magnetization, with the in-plane saturation fields ($H_s$) for the $Ce_{1.4}Y_{1.6}Fe_xO_{12}$ (x = 4.73, 4.89, 5.15, 5.47) series measured at 2 kOe. The saturation magnetization ($M_s$) of the $Ce_{1.4}Y_{1.6}Fe_xO_{12}$ (x = 4.73, 4.89, 5.15, 5.47) films were 145, 137, 133, and 117 emu/cm³, respectively. Above the compensation temperature, the saturation magnetization ($M_s$) of Ce:YIG is determined by the ferrimagnetic coupling between sublattices, expressed as $M_s = |M_d - M_a - M_c|$. Given that the tetrahedral (d) site hosts a larger population of $Fe^{3+}$ ions than the octahedral (a) site (in a 3:2 ratio), the net moment aligns with the d-sublattice. Our observation that $M_s$ decreased with increased iron content indicated that the excess iron predominantly occupied the octahedral (a) sites.

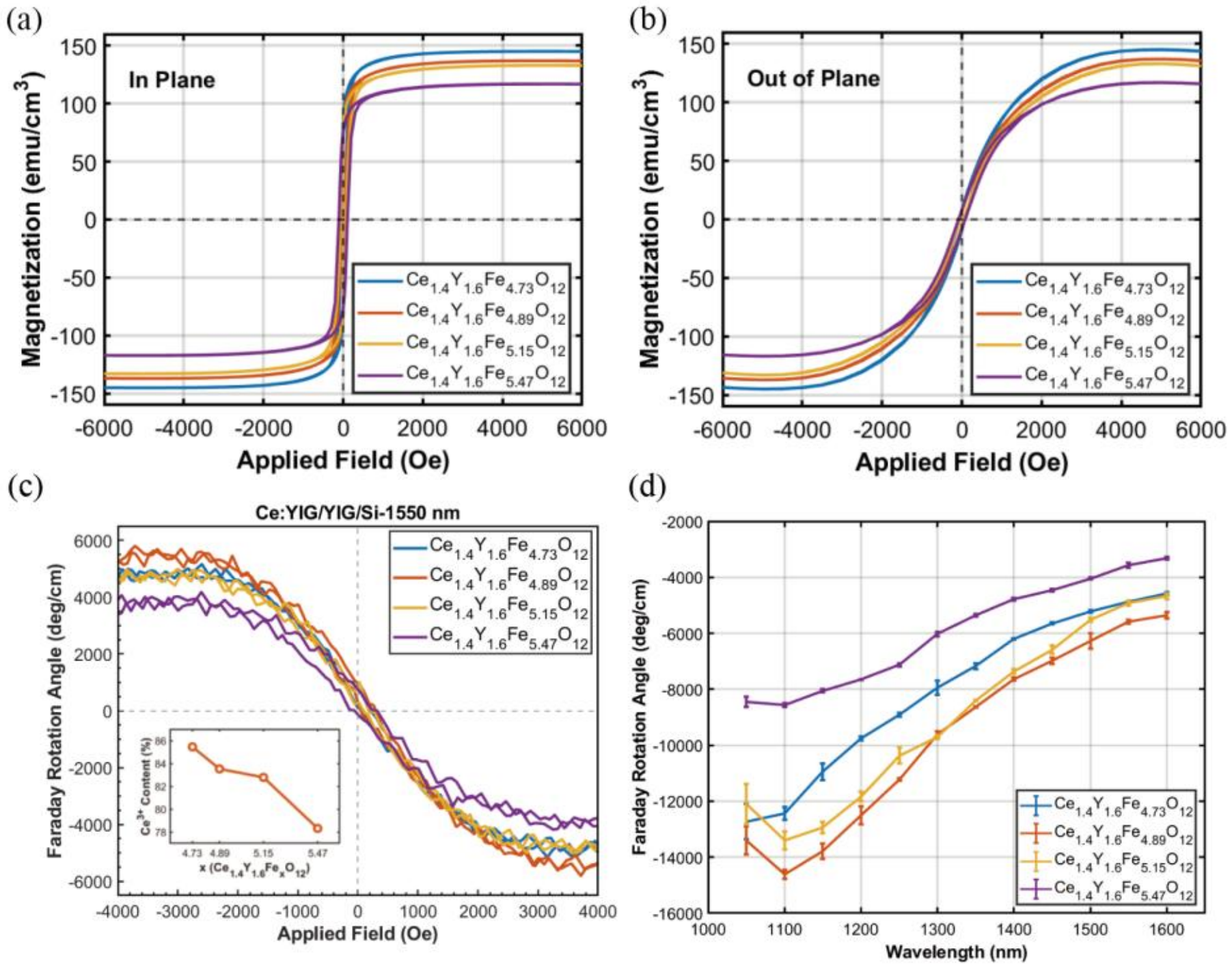

Fig. 3. Magnetic and magneto-optical properties of the Ce:YIG thin films. (a) In-plane and (b) out-of-plane magnetic hysteresis loops of the four thin films ($Ce_{1.4}Y_{1.6}Fe_xO_{12}$, x = 4.73, 4.89, 5.15, 5.47) measured by vibrating sample magnetometry (VSM). (c) Free space Faraday rotation hysteresis loops of the films in 1550 nm. (d) Magneto-optical spectra of the films measured across a wavelength range from 1050 to 1600 nm.

Figure 3(c) illustrated the free-space Faraday rotation (FR) hysteresis loops under an out-of-plane magnetic field. At 1550 nm wavelength, the nearly stoichiometric $Ce_{1.4}Y_{1.6}Fe_{4.89}O_{12}$ film exhibited the highest FR of -5523 ± 120 deg/cm. For the $Ce_{1.4}Y_{1.6}Fe_{4.73}O_{12}$ and $Ce_{1.4}Y_{1.6}Fe_{5.15}O_{12}$ films, the FR values decreased to approximately -4800 ± 125 deg/cm due to compositional deviations from stoichiometry. For the $Ce_{1.4}Y_{1.6}Fe_{5.47}O_{12}$ film, which significantly deviated from the ideal stoichiometric ratio, showed the minimum FR of only -3790 ± 123 deg/cm. The enhanced magneto-optical effect in Ce:YIG film primarily originates from the charge transfer transitions between the 4f electrons of $Ce^{3+}$ and 3d electrons of $Fe^{3+}$ [32-37]. XPS characterizations and subsequent peak fitting in Figure S1(b) revealed that the $Ce^{3+}$ concentrations in the $Ce_{1.4}Y_{1.6}Fe_xO_{12}$ (x = 4.73, 4.89, 5.15, 5.47) films were 85.45%, 83.52%, 82.78%, and 78.31%, respectively. The relative proportion of $Ce^{3+}$ ions monotonically increased with decreasing Fe concentrations. Relative to the ideal stoichiometry of $Ce_{1.4}Y_{1.6}Fe_{5.0}O_{12}$, the deposited $Ce_{1.4}Y_{1.6}Fe_xO_{12}$ (x = 4.73, 4.89, 5.15, and 5.47) films exhibited compositional deviations of -5.4%, -2.2%, 3.0%, and 9.4%, respectively. By extrapolating this compositional dependence, it could be deduced that a perfectly stoichiometric $Ce_{1.4}Y_{1.6}Fe_{5.0}O_{12}$ film would yield an optimal $Ce^{3+}$ proportion of approximately 86%.

Furthermore, the wavelength-dependent MO spectra of the four films across the range of 1050-1600 nm were shown in Figure 3(d). The variation trend of the absolute FR values remained consistent with the observations at 1550 nm. Notably, the peak response wavelength in the MO spectra shifted as a function of Fe content. This spectral evolution was attributed to the variation in the valence states of the constituent elements, which modified the electronic transitions responsible for the MO effect. Films with different stoichiometries alter the crystal field splitting energy of the $Fe^{3+}$ ions, which consequently causes the magneto-optical spectral peak to shift as a function of the Fe content [37, 38].

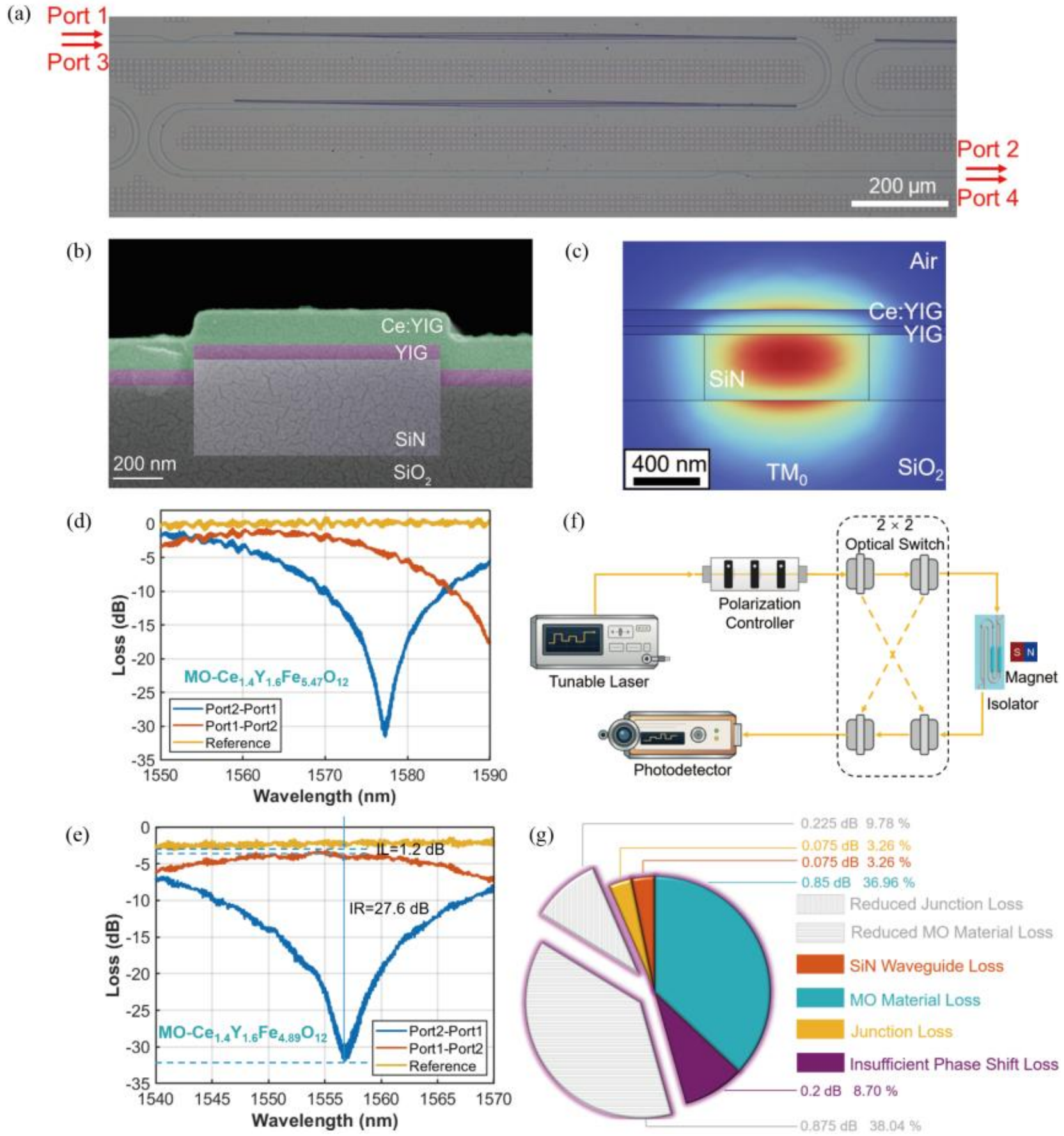

Fig. 4. Device architecture and nonreciprocal optical performance of the integrated MZI isolator. (a) Optical microscope image of the fabricated MZI device. (b) Cross-sectional SEM image of the SiN waveguide clad with the Ce:YIG/YIG magneto-optical film. (c) Simulated cross-sectional electric field distribution of the fundamental $TM_0$ mode. Transmission spectra of the MZI devices integrated with (d) $Ce_{1.4}Y_{1.6}Fe_{5.47}O_{12}$ and (e) $Ce_{1.4}Y_{1.6}Fe_{4.89}O_{12}$ MO claddings. (f) Schematic diagram of the experimental characterization setup. (g) Quantitative loss breakdown of the optimized low-loss device.

To evaluate the device performance, we characterized the on-chip Mach-Zehnder interferometer (MZI) optical isolators based on the silicon nitride (SiN) waveguide platform, as summarized in Figure 4. Figure 4(a) showed the optical microscope image of a TM mode optical isolator based on a SiN. The device design assumed a MO cladding composed of Ce:YIG/YIG. Simulations conducted via COMSOL revealed a nonreciprocal phase shift $\Delta\beta$ under these conditions. Accordingly, to accumulate the requisite $\pi/2$ phase shift in the MO section, the exposed window length was determined to be 621 μm. Furthermore, given that the fundamental propagation constant $\beta$ of the SiN waveguide, the RPS was engineered to introduce a phase shift. This yielded an RPS length of 24.88 μm.

Figure 4(b) showed the cross-sectional SEM image of the MO/SiN waveguide. The planar Ce:YIG (130 nm)/YIG (55 nm) on SiN formed a strip-loaded structure, avoiding unwanted sidewall MO deposition and preserving the TM-mode FoM. In this configuration, the SiN waveguide thickness was fixed at 400 nm. To ensure single-mode TM operation and suppress the excitation of higher-order modes, the SiN waveguide was designed with a thickness of 400 nm and a width of 1 μm to support fundamental $TM_0$-mode operation while maintaining optical confinement and interaction with the overlying Ce:YIG magneto-optical cladding. This geometry was selected to balance mode confinement and magneto-optical interaction in the MZI isolator. Finite-element method (FEM) simulations shown in Figure 4(c) illustrated the simulated $E_y$ field distribution of the fundamental $TM_0$ mode. Dictated by electromagnetic boundary conditions, the modal intensity exhibited a pronounced discontinuity at the material interfaces along the y-direction. Owing to the ultra-low loss of the SiN platform, the waveguide loss contributed by the SiN core within the MO waveguide section was considered negligible.

Figures 4(d) and 4(e) plotted the transmission spectra of the TM-mode SiN optical isolators integrated with the $Ce_{1.4}Y_{1.6}Fe_{5.47}O_{12}$ and $Ce_{1.4}Y_{1.6}Fe_{4.89}O_{12}$ MO claddings, respectively. The $Ce_{1.4}Y_{1.6}Fe_{5.47}O_{12}$ device exhibited severely degraded performance due to a prominent phase shift deficiency stemming from its inadequate Faraday rotation. In contrast, the device employing the nearly stoichiometric $Ce_{1.4}Y_{1.6}Fe_{4.89}O_{12}$ film demonstrated exceptional characteristics. At 1556 nm wavelength, the device achieved a maximum isolation ratio of 27.6 dB with a record-low insertion loss of 1.2 dB. Furthermore, while FEM simulations predicted a $0.5\pi$ nonreciprocal phase shift under an assumed Faraday rotation (FR) of -7000 deg/cm, the experimentally extracted phase shift was $0.413\pi$. This discrepancy indicated an actual FR of -5787 deg/cm for the deposited $Ce_{1.4}Y_{1.6}Fe_{4.89}O_{12}$ film. The figure of merit (FoM) of the isolator is defined by the following equation:

$$FoM = \frac{\Delta\beta}{\alpha} \quad (1)$$

$$\Delta\beta = \frac{2\beta}{\omega\varepsilon_0 N} \iint \frac{\gamma}{\mathrm{n}_0^4} H_x \partial_y H_x dxdy \quad (2)$$

$$\alpha = \Gamma_{Ce:YIG}\alpha_{Ce:YIG} + \Gamma_{YIG}\alpha_{YIG} + \Gamma_{SiN}\alpha_{SiN} \quad (3)$$

here, $\Delta\beta$ and $\beta$ denote the NRPS and the propagation constant of the fundamental TM mode, respectively. $\omega$ is the angular frequency, $\varepsilon_0$ is the vacuum permittivity, and N represents the z-directed power flux. For the MO material, $\gamma$ and $n_0$ correspond to the off-diagonal permittivity tensor component and the refractive index, respectively. Furthermore, $H_x$ is the magnetic field along the x-axis, while $\Gamma_{Ce:YIG}$, $\Gamma_{YIG}$ and $\Gamma_{SiN}$ represent the optical confinement factors within the Ce:YIG, YIG and SiN waveguide cores, respectively.

The testing system for the spectra was shown in Figure 4(f). Device characterization was performed using a polarization-maintaining (PM) end-fire coupling setup. A 2000 Oe in-plane magnetic field, perpendicular to the light propagation, was applied to saturate the MO cladding. Continuous-wave light (1520–1610 nm, 8 dBm) from a tunable laser was polarization-controlled and coupled into the device via lensed PM fibers. The transmission spectra were recorded by an optical sensor with a 2-pm resolution. For reverse transmission measurements, the magnetic field remained static while the input and output ports were swapped via a 2×2 optical switch.

Compared to the previous state-of-the-art on-chip MZI isolators featuring a 2.3 dB insertion loss, our optimized device achieved a record-low insertion loss of 1.2 dB, representing a reduction of 47.8%. Figure 4(g) presented a quantitative loss breakdown for the $Ce_{1.4}Y_{1.6}Fe_{4.89}O_{12}$ based isolator, normalized against the previous 2.3 dB baseline. With the waveguide architecture unchanged and the fabrication process optimized, both the SiN waveguide loss and junction loss are 0.075 dB (each accounting for 3.26% of the baseline). Junction loss was reduced by 0.225 dB due to the optimization of the window-opening process. The most significant contribution was the 0.875 dB loss reduction (38.04%), attributed to the eliminated secondary phase precipitation of the $Ce_{1.4}Y_{1.6}Fe_{4.89}O_{12}$ MO cladding. Notably, a 17.3% deviation between the actual and designed FR led to an insufficient nonreciprocal phase shift and imperfect optical interference, thereby introducing a 0.2 dB loss (8.70%). The remaining 0.85 dB (36.96%) originated from the intrinsic material loss of the $Ce_{1.4}Y_{1.6}Fe_{4.89}O_{12}$ film. Consequently, the 621 μm MO interaction length provided approximately 80% of the target nonreciprocal phase shift, resulting in imperfect interference and an estimated phase-mismatch penalty of 0.2 dB. To achieve the full target phase shift, the MO interaction length would need to be increased to approximately 776 μm, assuming a constant nonreciprocal phase shift per unit length. Accordingly, the estimated MO propagation loss would increase from 0.85 dB to 1.06 dB. Including the combined 0.15 dB SiN waveguide and junction losses, the total insertion loss would be approximately 1.21 dB, assuming the phase-mismatch penalty is eliminated and other loss contributions remain unchanged. This analysis indicates that simply increasing the MO interaction length does not necessarily reduce the total insertion loss, highlighting the need to further improve the magneto-optical figure of merit and optimize the phase-shifter design. The FoM of the $Ce_{1.4}Y_{1.6}Fe_{4.89}O_{12}$ film was calculated to be 72.7 deg/dB.

To evaluate the device performance, Table 1 benchmarked the proposed isolator against several MO isolators demonstrated on silicon-on-insulator (SOI) and SiN platforms. The comparative analysis revealed that our MZI-type MO isolator exhibited the lowest insertion loss reported to date. Moving forward, there remains significant potential to further drive down the insertion loss through continued material optimization. Although the elimination of defects has yielded a 54.5% reduction in the optical loss of the MO material, a discrepancy still exists when compared to the intrinsic loss of bulk Ce:YIG single crystals [39-41]. Although not quantified here, the polycrystalline YIG seed layer contributed to MO-related loss, which could be further reduced by employing more transparent garnet-based seed layers.

**Table 1. Comparison of Silicon-Integrated MZI MO Isolator Performance.**

| Device Type | Insertion Loss (dB) | Isolation Ratio (dB) | Monolithic /Bonding | Mode (TE/TM) | Ref. |
|---|---|---|---|---|---|
| SiN MZI | 1.2 | 27.6 | Monolithic | TM | This work |
| SiN MZI | 2.3 | 32 | Monolithic | TM | [16] |
| SiN MZI | 3 | 30 | Monolithic | TE | [16] |
| SOI MZI | 5 | 30 | Monolithic | TM | [17] |
| SOI MZI | 8 | 25 | Bonding | TM | [19] |
| SOI MZI | 8.2 | 12 | Bonding | TM | [20] |
| SOI MZI | 4.6 | 20 | Bonding | TE | [20] |

## 4. Conclusion

In conclusion, we observed that Ce:YIG film stoichiometry strongly influences the nanoscale phase purity and the FoM of Ce:YIG. By compensating for the preferential ablation of Fe during PLD using an iron-deficient target, we successfully synthesized nearly stoichiometric $Ce_{1.4}Y_{1.6}Fe_{4.89}O_{12}$ film. With significantly reduced Fe- and Ce-rich precipitates, yielding an exceptional FoM of 72.7 deg/dB. Consequently, the fabricated SiN-based MZI isolator achieved a record-low insertion loss of 1.2 dB alongside a maximum isolation ratio of 27.6 dB at the 1556 nm wavelength. These results demonstrate monolithically integrated Ce:YIG thin films and nonreciprocal photonic devices can achieve low insertion loss, paving the way for optical and data communication applications.

**Funding.**

National Natural Science Foundation of China (NSFC) (Grant Nos. 52650003，U22A20148 and 52473292) and Sichuan Provincial Science and Technology Department (Grant Nos. 2025ZYD0001 and 2025ZNSFSC0040).

**Disclosures.**

The authors declare no conflicts of interest.

See Supplement 1 for supporting content.

## 5. References

1. S. Ma, et al., "Magneto-optics enhances in-memory computing," Nature Photonics **19**, 9–10 (2025).
2. W. Tian, et al., "Photonic integrated circuits: research advances and challenges in interconnection and packaging technologies," in *Photonics*, (MDPI, 2025), 821.
3. J. Lapointe, et al., "Passive broadband Faraday isolator for hybrid integration to photonic circuits without lens and external magnet," Nature Photonics **19**, 248–257 (2025).
4. L. Fan, et al., "An all-silicon passive optical diode," Science **335**, 447–450 (2012).
5. K. Han, et al., "Integrated broadband optical isolator via dynamic rotating destructive interference," arXiv preprint arXiv:2509.02866 (2025).
6. S. Hua, et al., "Demonstration of a chip-based optical isolator with parametric amplification," Nature communications **7**, 13657 (2016).
7. Y. Shi, et al., "Limitations of nonlinear optical isolators due to dynamic reciprocity," Nature photonics **9**, 388–392 (2015).
8. G. Chaudhary and Y. Jeong, "Spatiotemporal modulated multi-functional Non-reciprocal filtering phase shifter with continuously tunable frequency and isolator functionalities," AEU-International Journal of Electronics and Communications, 156192 (2026).
9. C. Devitt, et al., "MEMS Switch Enabled Spatiotemporally Modulated Isolators," IEEE Transactions on Circuits and Systems I: Regular Papers (2025).
10. C. Doerr, et al., "Silicon photonics broadband modulation-based isolator," Optics express **22**, 4493–4498 (2014).
11. K. Chao, et al., "Integrated TE optical isolator based on magneto-optical perturbation in coupled waveguides," Scientific reports **15**, 20381 (2025).
12. Z. Wei, et al., "Recent development of magneto-optical thin films and integrated photonic devices," Journal of Materials Chemistry C **13**, 12628–12649 (2025).
13. L. Bi, et al., "On-chip optical isolation in monolithically integrated non-reciprocal optical resonators," Nature Photonics **5**, 758–762 (2011).
14. Q. Du, et al., "Monolithic On-chip Magneto-optical Isolator with 3 dB Insertion Loss and 40 dB Isolation Ratio," Acs Photonics **5**, 5010–5016 (2018).
15. T. Fakhrul, et al., "Magneto-optical Bi:YIG films with high figure of merit for nonreciprocal photonics," Advanced Optical Materials **7**, 1900056 (2019).
16. W. Yan, et al., "Waveguide-integrated high-performance magneto-optical isolators and circulators on silicon nitride platforms," Optica **7**, 1555–1562 (2020).
17. Y. Zhang, et al., "Monolithic integration of broadband optical isolators for polarization-diverse silicon photonics," Optica **6**, 473–478 (2019).
18. W. Yan, et al., "On-chip Integrated Magneto-Optical Nonreciprocal Photonic Devices," in *Integrated Photonics Research, Silicon and Nanophotonics*, (Optica Publishing Group, 2021), IM1B. 4.
19. S. Ghosh, et al., "Ce:YIG/Silicon-on-Insulator waveguide optical isolator realized by adhesive bonding," Optics express **20**, 1839–1848 (2012).
20. S. Liu, et al., "Silicon-based integrated polarization-independent magneto-optical isolator," Optica **10**, 373–378 (2023).
21. Y. Shoji, et al., "Silicon Mach–Zehnder interferometer optical isolator having 8 nm bandwidth for over 20 dB isolation," Japanese Journal of Applied Physics **53**, 022202 (2014).

22. I. Thorlabs, "Optical Isolators" (Thorlabs, Inc.), retrieved 2026-07-07, https://www.thorlabs.com/optical-isolators.
23. S. Higuchi, et al., "Magneto-optical properties of cerium-substituted yttrium iron garnet single crystals grown by traveling solvent floating zone method," Japanese journal of applied physics **38**, 4122 (1999).
24. M. Li, et al., "Design of a broadband and low loss TM magneto-optical waveguide isolator based on lithium niobate on insulator," IEEE Photonics Journal **13**, 1–11 (2021).
25. L. Liu, et al., "Integrated Low-Loss, High-Isolation, and Broadband Magneto-Optical Isolator with TE-Mode Input," Micromachines **16**, 315 (2025).
26. G. Portela, et al., "Novel compact magnetless isolator based on a magneto-optical garnet material," Optics & Laser Technology **157**, 108638 (2023).
27. M. C. Onbasli, et al., "Optical and magneto-optical behavior of cerium yttrium iron garnet thin films at wavelengths of 200–1770 nm," Scientific reports **6**, 23640 (2016).
28. Y. Yang, et al., "Europium-substituted cerium iron garnet thin films for silicon-integrated nonreciprocal photonic device applications," APL Materials **13**(2025).
29. C. B. Arnold and M. J. Aziz, "Stoichiometry issues in pulsed-laser deposition of alloys grown from multicomponent targets," Applied Physics A **69**, S23–S27 (1999).
30. T. C. Droubay, et al., "Nonstoichiometric material transfer in the pulsed laser deposition of $LaAlO_3$," Applied Physics Letters **97**(2010).
31. C. Yu, et al., "Stoichiometry, phase, and texture evolution in PLD-grown hexagonal barium ferrite films as a function of laser process parameters," Journal of Alloys and Compounds **814**, 152301 (2020).
32. 黄强, et al., "掺铈钇铁石榴石磁光薄膜光吸收机理研究," 磁性材料及器件 **29**, 40﹣44 (1998).
33. M. Kučera, "Magneto-optics of $Ce^{3+}$ doped garnets," Journal of magnetism and magnetic materials **101**, 242–244 (1991).
34. M. Kucera, et al., "Faraday rotation and MCD in Ce doped yig," Solid state communications **69**, 1117–1121 (1989).
35. X. Liang, et al., "First principles calculation on the magnetic, optical properties and oxygen vacancy effect of $Ce_xY_{3-x}Fe_5O_{12}$," Applied Physics Letters **106**(2015).
36. G. Scott, et al., "Absorption spectra of $Y_3Fe_5O_{12}$ (YIG) and $Y_3Ga_5O_{12}$: $Fe^{3+}$," Physical Review B **10**, 971 (1974).
37. H. B. Vasili, et al., "Direct observation of multivalent states and 4f→3d charge transfer in Ce-doped yttrium iron garnet thin films," Phys. Rev. B (2017).
38. M.-Y. Chern, et al., "Red shift of Faraday rotation in thin films of completely bismuth-substituted iron garnet $Bi_3Fe_5O_{12}$," Japanese Journal of Applied Physics **38**, 6687 (1999).
39. S. Higuchi, et al., "Magnetooptical properties of cerium-substituted yttrium iron garnet single crystals for magnetic-field sensor," Sensors and Actuators A: Physical **105**, 293–296 (2003).
40. A. Ikesue, et al., "Giant Faraday rotation in heavily ce-doped YIG bulk ceramics," Journal of the European Ceramic Society **40**, 6073–6078 (2020).
41. W. Jin, et al., "Crystal growth and characterization of $Ce_xY_{3-x}Fe_5O_{12}$ single crystal by optical floating zone method," Physica B: Condensed Matter **588**, 412168 (2020).